\documentclass[useAMS,usenatbib,prd,amsmath,amssymb,superscriptaddress]{mn2e}

\usepackage{graphics}
\usepackage{rotating}
\usepackage{epsfig}
\usepackage{color}

\newcommand{\bw}{\Delta\nu}

\DeclareGraphicsExtensions{.eps}

\title{Galaxy redshift surveys selected by neutral hydrogen using the Five-hundred metre Aperture Spherical Telescope}
\author[A. R. Duffy et al.]
{Alan R. Duffy,$^{1,2}$ Richard A. Battye,$^1$ Rod D. Davies,$^1$
\newauthor
 Adam Moss,$^3$ Peter N. Wilkinson$^1$ \\ 
$^1$Jodrell Bank Observatory, School of Physics and Astronomy, University of Manchester, Macclesfield, Cheshire SK11 9DL, U.K.\\
$^2$Leiden Observatory, Leiden University, PO Box 9513, 2300 RA Leiden, The Netherlands\\
$^3$Department of Physics and Astronomy, University of British Columbia, 6224 Agricultural Road, Vancouver, BC, V6T 1Z1, Canada}
\begin{document}

\date{12/07/2007}

\pagerange{\pageref{firstpage}--\pageref{lastpage}} \pubyear{2007}

\maketitle

\label{firstpage}

\begin{abstract}
We discuss the possibility of performing a substantial spectroscopic galaxy redshift survey selected via the 21cm emission from neutral hydrogen using the Five-hundred metre Aperture Spherical Telescope (FAST) to be built in China. We consider issues related to the estimation of the source counts and optimizations of the survey, and discuss the constraints on cosmological models that such a survey could provide. We find that a survey taking around 2 years could detect $\sim 10^{7}$ galaxies with an average redshift of $\sim 0.15$ making the survey complementary to those already carried out at optical wavelengths. These conservative estimates have used the $z=0$ HI mass function and have ignored the possibility of evolution. The results could be used to constrain $\Gamma=\Omega_{\rm m}h$ to 5\% and the spectral index,  $n_{\rm s}$, to 7\% independent of CMB data. If we also use simulated power spectra from the Planck satellite, one can constrain $w$ to be within 5\% of -1.
\end{abstract}


\section{Introduction}
\label{Introduction}

Galaxy redshift surveys have played a significant role in constraining the cosmological model, with the largest surveys to date having been performed by the 2dFGRS\footnote{2dF homepage: www.aao.gov.au/2dF} and SDSS\footnote{SDSS homepage: www.sdss.org} teams using optical techniques. 
Low redshift samples have been used for some time to derive constraints on cosmological parameter combinations $\Gamma$ and $f_{\rm b}=\Omega_{\rm b}/\Omega_{\rm m}$, where $\Omega_{\rm m}$ and $\Omega_{\rm b}$ are the total matter and baryon densities defined relative to critical, and $h=H_0/(100{\rm km}\,s^{-1}\,{\rm Mpc}^{-1})$, as well as the spectral index of the density fluctuations, $n_{\rm s}$, and neutrino densities~\citep{Tegmark2004a, Tegmark2004b, Percival, Cole}.
More recently the SDSS team have used a more sparsely sampled, but larger volume survey of luminous red galaxies (LRGs) to measure the expected baryon acoustic peak on scales of $\sim 100h^{-1}{\rm Mpc}$~\citep{SDSS1}.
Since the acoustic scale is a standard ruler, this provides a distance measurement allowing further constraints to be placed on cosmological parameters and, in particular, those which parametrize the properties of the dark energy.

A large number of projects have been proposed recently to develop these ideas and to perform large photometric surveys in the optical (and possibly the infra-red) waveband to find $\sim 10^{8}$ galaxies out to $z \sim 1$ (Dark Energy Survey; DES\footnote{DES homepage: www.darkenergysurvey.org}), as well as deeper spectroscopic surveys to find $\sim 10^{5}$ galaxies at $z \sim 3$~\citep{Blake}. Such surveys will have much larger search volumes than those presently available, allowing for significantly more accurate measurements of the power spectrum and hence of the acoustic scale at a variety of redshifts. This could lead to very tight constraints on the properties of the dark energy.

Another promising technique is to use the 21cm emission from neutral hydrogen (HI) as a tracer of the galaxies allowing for a spectroscopic galaxy redshift survey~\citep{AR}. This will require the development of telescopes with much higher survey speed, and hence a combination of larger collecting area and field-of-view (FoV), than those available at present. The current state-of-the-art is a shallow all-sky survey; the combined HIPASS~\citep{hipass} and HIJASS~\citep{hijass} survey, which has found $\sim 10^4$ galaxies. 
Fortunately, a world-wide technology development programme is underway with the ultimate aim of building a Square Kilometre Array (SKA\footnote{SKA homepage: www.skatelescope.org/documents}) with collecting area $\sim 10^{6}\,{\rm m}^2$ and a FoV much greater than $1\,{\rm deg}^2$, with these specifications the SKA surveys may detect $\sim 10^{9}$ galaxies.

Before the full SKA comes into operation ($\sim 2020$) another large radio telescope will have been constructed which also has great potential for HI surveys. The Five-hundred-metre Aperture Spherical Telescope (FAST) will be an Arecibo-like telescope with a significantly larger aperture ($\sim 500\,{\rm m}$) and FoV~\citep{KARST}. The aim of this paper is to discuss the scientific potential of a low-redshift HI survey of galaxies as a FAST key science programme. We will show that, if a focal plane array with 100 instantaneous beams can be developed, $\sim 10^{7}$ galaxies can be detected within a realistic survey period ($\sim 2$ years) and that the low-redshift galaxy power spectrum can be measured much more accurately than is currently possible with optical surveys. Since hydrogen is the fundamental baryonic constituent of the universe such a survey would provide an important view of the large-scale structure of the universe with different biases and be an important stepping stone to the larger HI galaxy surveys which will become possible with the SKA.

In the next section we will briefly discuss some relevant observational parameters of  FAST and then in subsequent sections, we will estimate the number counts, make forecasts for errors on the power spectrum and constraints on cosmological parameters that could be possible with such a survey.

\section{Five-hundred-metre Aperture Spherical Telescope (FAST)}

The FAST telescope is funded and when completed in around 2012-2013 it will be the largest single dish telescope in the world.  The expected specifications of the telescope~\citep{KARST} relevant to our discussion here are:
\begin{itemize}
\item The prime-focus  feed illuminates an area corresponding to a parabolic dish of diameter $\sim 300 {\rm m}$ with aperture efficiency of $70\%$ giving $A_{\rm eff}=50000 {\rm m}^{2}$,
\item  At 21cm wavelength the beam width (FWHM) $\theta \approx 3$ arcmin; the effective pixel (equating the volume of a cube of height unity to the integral under a gaussian beam with the same height) has side 1.064$\theta$ and area 1.133$\theta^2$.
\item The maximum observable zenith angle is of order $40^{\circ}$ allowing the observation of $\approx 50 \%$ of the full sky without significant degradation of performance. Henceforth, the `full sky' is defined to be the maximum amount of sky accessible to the telescope i.e.  $\approx 20000 {\rm deg}^{2}$.
\item One of the principal receiver systems will be a focal plane array covering a frequency range around 1.4 GHz (the 21-cm HI line) which can produce  $n_{\rm B}$ instantaneous beams each with a system temperature $T_{\rm sys}=25{\rm K}$ (hence $A_{\rm eff}/T_{\rm sys}\approx 2000\,{\rm m}^2\,K^{-1}$ per beam) and dual polarization capability. 
\end{itemize}
The expected thermal noise for a dual polarization single beam can be computed using 
\begin{eqnarray}\label{eq:flux noise} \sigma_{\rm noise}=\sqrt{2}\frac{kT_{\rm sys}}{A_{\rm eff}}\frac{1}{\sqrt{\bw t}}, \end{eqnarray}
for an observing time of $t$ and a frequency bandwidth of $\bw$, where $k=1380\,{\rm Jy}\,{\rm m}^2\,{\rm K}^{-1}$ is the Boltzmann constant. If we assume a bandwidth of $\Delta\nu=1\,{\rm MHz}$ which corresponds to a velocity linewidth of $\approx 200\,{\rm km}\,s^{-1}$ at 21cm then the instantaneous sensitivity of each beam of the FAST system will be $\sigma_{\rm inst}\approx 1\,{\rm mJy}\,s^{1/2}$.

The focal ratio of the telescope ($f/D=0.47$) restricts $n_{\rm B}$ to $\le 19$ for a conventional close-packed horn-based array, similar to that used for HIPASS and operated at the prime focus. Larger arrays require some of the horns to be far enough off-axis that aberration losses become unacceptable. A 19-beam array gives an instantaneous FoV of $\Omega_{\rm inst}\approx 200\,{\rm arcmin}^2$. We will use this 19-beam system as the fiducial benchmark for our basic calculations. A receiver array with $n_{\rm B}\sim 100$ may well be possible. This requirement could only be met with close-packed phased arrays of small antenna elements; these would not take up much more area than the 19-beam horn system. The required phased array technology is actively being developed by several groups working within the international Square Kilometre Array R\&D effort\footnote{SKA homepage: www.skatelescope.org/documents} and hence we can be confident that by the time FAST comes into operation in 2012-3 the 100-beam receiver capability will be available.  This would allow for $\Omega_{\rm inst}\approx (n_{\rm B}/19)200\,{\rm arcmin}^2$. Such a system will be at least twice as sensitive per beam as that on the Arecibo Telescope and have a survey speed ($\Omega_{\rm inst}(A_{\rm eff}/T_{\rm sys})^2$) which is at least $6(n_{\rm B}/19)$ times faster than the 7-beam system presently available there. FAST will also have more frequency flexibility due to the excellent RFI enviroment at the FAST site and will be able to observe twice the area of the sky than does the Arecibo Telescope.

\section{Estimating galaxy counts}

\subsection{Properties of the HI galaxy distribution}
\label{hidist}

The HI mass, $M_{\rm HI}$, of a galaxy at redshift $z$ is given in terms of the observed flux, $S$, and line width, $\Delta V_{\rm o}$, by~\citep{Roberts}
\begin{eqnarray}
\label{eq:mass limit} 
\frac{M_{\rm HI}}{M_{\odot}} = \frac{2.35 \times 10^{5}}{1+z} \left( \frac{d_{\rm L}(z)}{\rm Mpc} \right)^{2} \left(\frac{S}{\rm Jy} \right) \left(\frac{\Delta V_{\rm o}}{{\rm km}\,{\rm sec^{-1}}} \right), 
\end{eqnarray}
where $d_{\rm L}(z)$ is the luminosity distance to the galaxy. We note that this formula is usually quoted in terms of the Euclidean distance and without the $(1+z)^{-1}$; this is the correct formula in an FRW universe.

One can estimate the number of galaxies above some limiting mass $M_{\rm lim}(z)$ by computing
\begin{eqnarray}\label{eq:zeroth} N\left( M > M_{\rm lim},z\right) = \Delta\Omega \Delta z \frac{dV}{dzd\Omega} \int_{M_{\rm lim} \left( z \right)}^{\infty}\,\frac{dN}{dVdM}\,dM, \end{eqnarray}
where the  sky area covered is  $\Delta\Omega$ and the size of the redshift bin is $\Delta z$. $dV/dzd\Omega$ is the comoving volume element for the FRW universe and $dN/dVdM$ is the comoving number density of galaxies per unit mass which we shall assume can be approximated by a Schechter function 
\begin{eqnarray}
\label{eq:schechterfn}
{dN\over dVdM}={\theta^{*}\over M^*_{\rm HI}}\left({M\over M^*_{\rm HI}}\right)^\alpha\exp\left[-{M\over M^*_{\rm HI}}\right]\,.
\end{eqnarray}
We will, for the most part, assume no evolution in the HI mass function out to the moderate redshifts appropriate for a FAST survey and, therefore, use values of the parameters derived from the most recent survey of the local universe~\citep{hipass}. In particular, we will choose $\alpha = -1.37$, $\theta^{*}=1.42 \times 10^{-2} h^{3}{\rm Mpc}^{-3}$ and $M_{\rm HI}^{*}=10^{9.8}M_{\odot}$.

One can compute the average redshift of galaxies in the survey from $N(M>M_{\rm lim},z)$ by integrating appropriately over $z$, that is, 
\begin{eqnarray}
\langle z\rangle = {\int_0^{\infty}z\,N(M>M_{\rm lim},z)\,dz
\over \int_0^{\infty}N(M>M_{\rm lim},z)dz}\,.
\end{eqnarray}

In order to compute the limiting mass of the survey we will make a point source approximation, that is, we will assume that each galaxy detected will be smaller than the telescope beam, and hence, all the flux it emits will contribute to a single beam. This is likely to be a good approximation for most objects in the survey, but will break down for the closest, most massive objects. We expect those objects which violate this assumption to be the minority if we choose the survey strategy sensibly.

\begin{figure}
  \begin{center}
    \epsfysize=2in
    \epsfxsize=4in
    \epsfig{figure=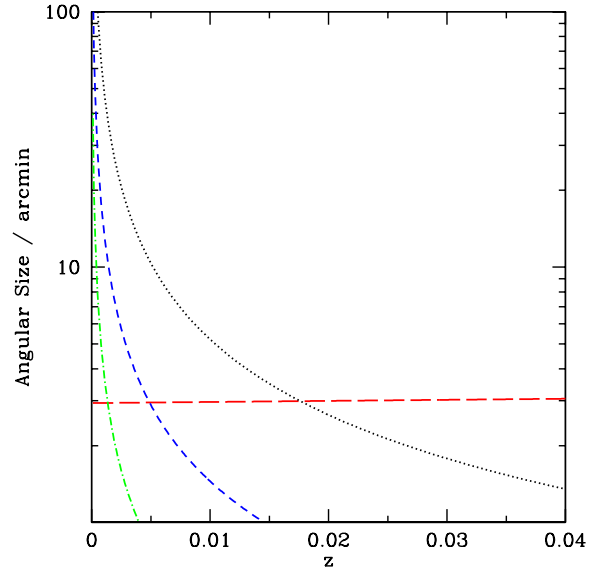, scale=0.4}
    \caption{The angular sizes (in arc minutes) for galaxies with HI masses $10^{10}M_{\odot}$, dotted, $10^{9}M_{\odot}$, dashed, and $10^{8}M_{\odot}$,  dot-dash. The (approximately) horizontal large-dashed line represents the FWHM beamsize of FAST which scales as $\lambda\propto 1+z$.}
    \label{fig:rda}
  \end{center}
\end{figure}

There is an empirically derived relation between the HI diameter, $D_{\rm HI}$, which is defined to be the region inside which the HI surface density is greater than $1M_{\odot}\,{\rm pc}^{-2}$, and the HI mass of a galaxy. This can be used to investigate the extent to which the point source approximation is valid. We will use~\citep{BR,VS}
\begin{eqnarray}\label{eq:size mass relation} \frac{D_{\rm HI}}{\rm kpc}=\left( \frac{M_{\rm HI}}{10^{6.8}M_{\odot}} \right)^{0.55}\,,
\end{eqnarray}
which has been converted to an angular scale using the angular diameter distance $d_{\rm A}(z)$ for the range of relevant HI masses in Fig.~\ref{fig:rda}.
We note immediately that at $z\approx 0.02$ massive galaxies containing $10^{10}M_{\odot}$ of HI will be resolved by the FAST beam (see Fig.~\ref{fig:rda}) but we postpone more detailed discussion of the validity of the point source assumptions to later in section B.

The beam area increases like $\lambda^{2}\propto (1+z)^2$. As a result if an optimal survey strategy, in which one uniformly tiles the $z=0$ slice, is chosen then slices of the survey at higher redshift receive extra exposure due to the fact that these slices will overlap. The effect has been considered in ref.~\citep{AR} where it was shown that the flux limit relevant to a particular redshift slice is reduced by $(1+z)^{-1}$. Hence, the flux limit for an observation, $S_{\rm lim}$, for a specific signal-to-noise ratio $(S/N)$ is given by
\begin{eqnarray}\label{eq:fluxlimit} S_{\rm lim}= (S/N) \frac{\sigma_{\rm noise}}{1+z}\,.\end{eqnarray}
A given object is detectable if its flux density in bandwidth $\Delta\nu=\nu\Delta V_o/c$ is greater than $S_{\rm lim}$. 

\begin{figure}
  \begin{center}
    \epsfysize=2in
    \epsfxsize=4in
    \epsfig{figure=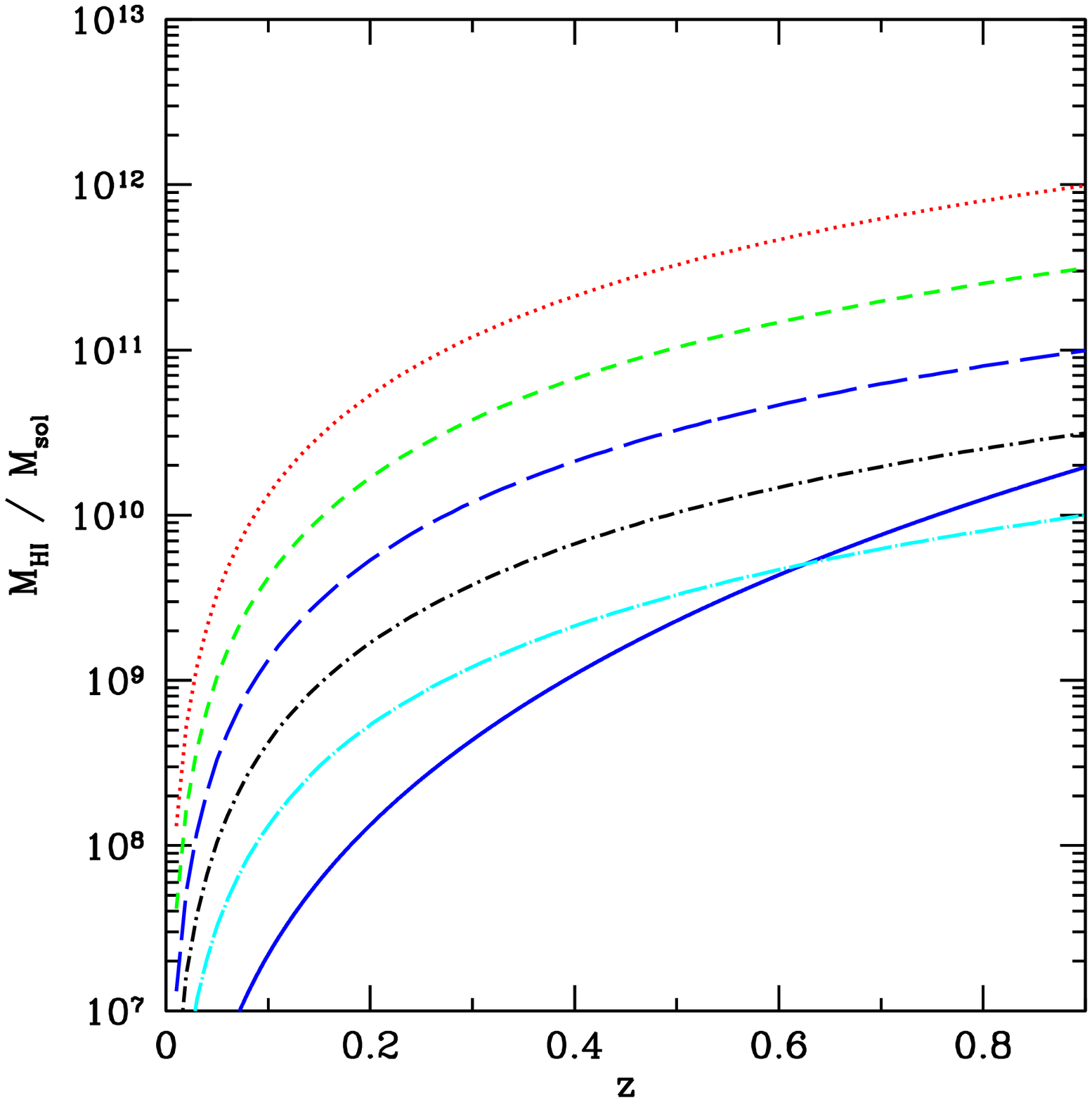, scale=0.4}
    \epsfig{figure=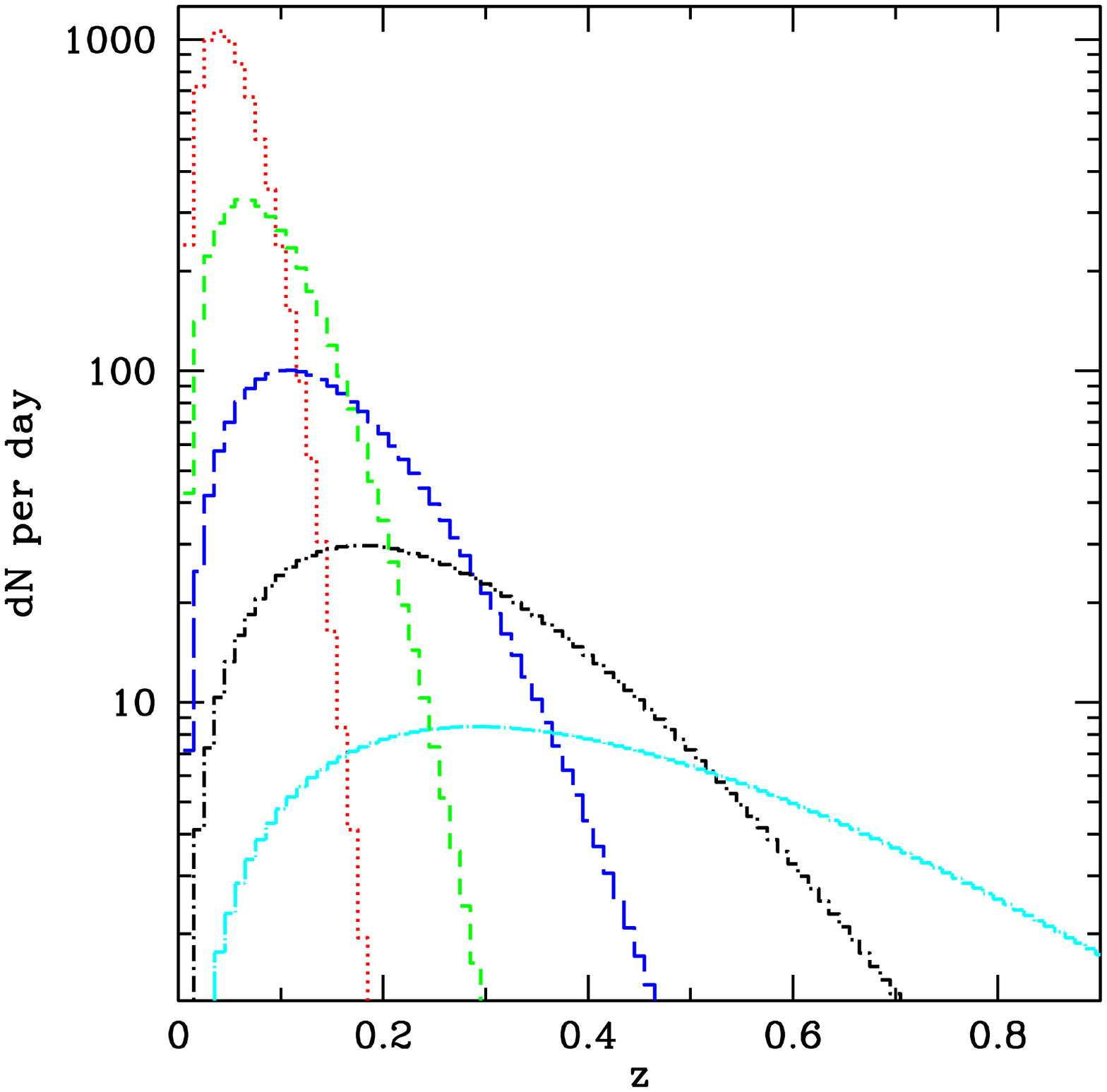, scale=0.4}
    \caption{a) the $S/N=4$ limiting mass for different integration times (in seconds) per pointing: $t_{\rm obs}=6$ (dotted), 60 (short-dash), 600 (long-dash), 6000 (dot-short dash), 60000 (dot-long dash). 
b) the predicted number counts in a day of observation (defined as 18 hours of on-source integration time) for each of the observation strategies for $n_{\rm B}=19$. The bin width is $\Delta z=0.01$. The total number of galaxies predicted in each case is summarized in Table~\ref{tab1}. The solid line (that is, the lowest) in panel a) represents the estimated HI mass in a typical volume enclosed by the beam area and the fiducial velocity width $200\,{\rm km}\,s^{-1}$. If this line is greater than a substantive fraction (say, 1/3) of the mass limit one would expect a confusion related increase in the observed noise.}
    \label{fig:simple}
  \end{center}
\end{figure}

\begin{center}
\begin{table}
\begin{tabular}{|c||c|c||} \hline
 $t_{\rm obs}/s$ &  $S_{\rm lim}/{\rm \mu Jy}$ &  $N_{4\sigma}(n_{\rm B}=19,t=1\,d)$ \\ \hline

6 & 1700 & 7000 \\ \hline

60 & 520 & 3800 \\ \hline

600 & 170 & 2000 \\ \hline

6000 & 50 & 1000 \\ \hline

60000 & 17 & 510 \\ \hline

\end{tabular}
\caption{\label{tab1} Predicted number counts with $S/N=4$  in one day of observation using a receiver system with $n_{\rm B}=19$ for a range of observation strategies. Counts expected for longer observations, or for larger numbers of beams, can be computed using a simple linear scaling. The limiting fluxes are computed using Eqn.~\ref{eq:fluxlimit} with $z=0$. All numbers have been rounded to 2 significant figures.}
\end{table}
\end{center}

\subsection{Simple estimates assuming a fixed observing bandwidth}
\label{simple}

A simple assumption, which should yield an order of magnitude estimate of the number of objects that one might expect to find in a survey, is that the line width of {\it all} the galaxies is $\Delta V_{\rm o}=200\,{\rm km}\,s^{-1}$. Under this assumption we have computed $M_{\rm lim}(z)$ and the number of galaxies one would expect to find in an average day of observation as a function of $t_{\rm obs}$, the amount of time spent observing each beam area. Allowing for calibration and telescope maintainance, we will use an on-source integration time of 18 hours to represent one day. The mass limit and number counts as a function of redshift (with bin width $\Delta z=0.01$) are presented in Fig.~\ref{fig:simple} for $S/N=4$ for $n_{\rm B}=19$. The probability of a spurious detection will be $\sim 6\times 10^{-5}$, assuming a Gaussian distribution for the noise. Since an all sky survey will contain $\sim 4\times 10^{9}$ independent volume elements (the number of independent beam areas multiplied by the number of indepndent depth slices) this would mean $\sim 2\times 10^{5}$ spurious detections. In such a survey, therefore, there will be around $1\%$ spurious detections (we will find that around $10^{7}$ galaxies can be detected). This will not significantly affect the measurement of the power spectrum discussed in the later sections of this paper.

One can immediately scale the number counts discussed in the previous paragraph to longer integration times and larger numbers of beams. If $t_{\rm survey}$ is the total time of the survey and the focal plane array has $n_{\rm B}$ beams then the number of galaxies detected at a particular signal-to-noise ratio is given by 
\begin{eqnarray} 
N(n_{\rm B},t_{\rm survey})= \left( \frac{n_{\rm B}}{19} \right) \left( \frac{t_{\rm survey}}{1\,d} \right) N(n_{\rm B}=19,t=1\,d)\,, 
\end{eqnarray}
although one has to be careful to make sure that $t_{\rm survey}<t_{\rm sky}$, the time taken to cover the entire sky available to FAST, which is given in terms of $t_{\rm obs}$ by 
\begin{eqnarray} \left( \frac{t_{\rm sky}}{1\,d} \right) = 5.5\left( \frac{19}{n_{\rm B}} \right) \left( \frac{t_{\rm obs}}{1s} \right)\,,
\end{eqnarray}
that is, the full $20000\,{\rm deg}^2$ will be covered in $5.5$ days for a system with 19 beams if one spent $1\,s$ per pointing. We remind the reader that we have assumed that 1 day corresponds to 18 hours of on-source integration time.

Under the assumptions of this calculation, we deduce, for example, that a survey of the whole sky available to FAST with a noise level of $\sigma_{\rm noise}\approx 40\,\mu{\rm Jy}$ per beam is possible by performing $600\,s$ integrations  in around 9 years if $n_{\rm B}=19$, but that this would take less than 2 years with $n_{\rm B}\sim 100$. Such a survey would find around $7\times 10^{6}$ galaxies with $S/N=4$ with an average redshift $\langle z\rangle\approx 0.15$. In subsequent sections, we will improve the validity of the calculation by modelling additional important effects. However, we shall see that this simple calculation holds true as a first order estimate. It is already clear that, if such a cosmologically important redshift survey is to be completed in an acceptable amount of time, a focal plane array with $n_{\rm B}\sim 100$ is essential.

We can check {\it a posteriori} that the point source approximation that we have made is valid by examining the limiting mass and expected number counts at very low redshift. We note first that the number counts tend to zero as $z$ tends to zero since the volume element is $\propto z^2$. Thus, even in the case of $t_{\rm obs}=6\,s$, there are very few objects with $z<0.02$, the point at which $M_{\rm HI}=10^{10}\,M_{\odot}$ galaxies are resolved by FAST. The percentage of low redshift objects will decrease as $t_{\rm obs}$ increases. We estimate that the percentage of galaxies that would be significantly resolved for $t_{\rm obs}=6\,s$ is less than $10\%$ and for $t_{\rm obs}>60\,s$ the effect will be negligible.

In the most extreme circumstances, one might also be sensitive to the effects of confusion even in a spectroscopic survey since the FAST beamsize is likely to be much larger than the size of the galaxies we are trying to detect. One can attempt to quantify the expected amount of neutral hydrogen in volume of the universe enclosed by a Gaussian beam with area $\Omega_{\rm B}=\pi\theta_{\rm FWHM}^2/(4\log2)$ and the fiducial linewidth $\Delta V_{\rm o}=200\,{\rm km}\,s^{-1}$. This is given by 
\begin{eqnarray}
\label{eq:confusion} 
\langle M_{\rm HI}\rangle=\rho_{\rm HI}(z)(1+z)\frac{\Delta V_{\rm o}}{c} \Omega_{\rm B}\frac{dV}{dzd\Omega}\,,
 \end{eqnarray}
where conservatively we assume that $\rho_{\rm HI}(z)=\rho_{\rm HI}(0)(1+z)^{3}$. We take the local HI density to be $\Omega_{\rm HI}=2.6 \times 10^{-4}{h}^{-1}$ as found in the HIPASS survey~\citep{hipass}, that is, $\rho_{\rm HI}(0)\approx 7.2\times 10^{7}\,h\,M_{\odot}\,{\rm Mpc}^{-3}$. The computed value of $\langle M_{\rm HI}\rangle(z)$ is compared to the mass limits for various values of $t_{\rm obs}$ in Fig.~\ref{fig:simple}. It is clear from this that the effects of confusion are only significant for large values of $t_{\rm obs}$; one might imagine that if $\langle M_{\rm HI}\rangle > M_{\rm lim}/3$ then confusion could lead to an increase in the effective noise making it more difficult to find galaxies.

From the above discussion it seems sensible to consider values of $60\,s<t_{\rm obs}<6000\,s$, in order to avoid any possible corrections to the statistics due to resolving galaxies at low $z$ and confusion at high $z$.

\subsection{Mass and inclination dependent effects in the linewidth}
\label{inclination}

In this section we attempt to improve our calculation of the number counts by taking into account the fact that not all galaxies have the same observed linewidth $\Delta V_{\rm o}$. This can happen for two reasons: first the intrinsic linewidth is expected to depend on the mass for virialized systems, and secondly not all galaxies will be observed edge-on. If they are observed face-on, then it is the motion of HI perpendicular to the disk which sets the linewidth and not rotation.

The intrinsic linewidth of a galaxy, corrected for broadening, has been shown empirically to be related to the HI mass by~\citep{BriggsRao:1993, hijass}
\begin{eqnarray}
\label{eq:velocity-mass relation}
\frac{\Delta V_{\rm e}}{420\,{\rm km}\,s^{-1}}=\left( \frac{M_{\rm HI}}{10^{10}M_{\odot}} \right)^{0.3}\,, \end{eqnarray} 
although we note that this relation shows a large dispersion, especially for dwarf galaxies. 

The observed linewidth of a galaxy, $\Delta V_\theta$, which subtends an angle $\theta$ between its spin axis and the line-of-sight can be computed using the Tully-Fouque rotation scheme~\citep{TFq}
\begin{eqnarray}\label{eq:TFq}
\lefteqn{({\Delta V_{\rm e} \sin(\theta)})^{2} = } \nonumber \\ 
& & (\Delta V_{\theta})^{2} + (\Delta V_{\rm t})^{2} - \nonumber \\
& & 2{\Delta V_{\theta}}{\Delta V_{\rm t}}\left( 1- e^{- \left(\frac{{\Delta V_{\theta}}}{\Delta V_{\rm c}}\right)^{2}} \right) - 2(\Delta V_{\rm t})^{2}e^{- \left(\frac{{\Delta V_{\theta}}}{\Delta V_{\rm c}}\right)^{2}}\,.
\end{eqnarray}
$\Delta V_{\rm c} = 120\,{\rm km}\,s^{-1}$ represents an intermediate transition between the small galaxies with Gaussian HI profiles in which the velocity contributions add quadratically and giant galaxies with a `boxy' profile reproduced by the linear addition of the velocity terms. $\Delta V_{\rm t}\approx 20\,{\rm km}\,s^{-1}$ is the velocity width due to  random motions in the disk~\citep{Rhee96,VS}. 

With this definition of $\theta$, $\theta=0$ corresponds to face-on and $\theta=\pi/2$ to edge-on. In cases where $\Delta V_\theta>>\Delta V_{\rm c}$, one can see that $\Delta V_\theta=\Delta V_{\rm t}+\Delta V_{\rm e}\sin\theta$. For $\theta=0$, one finds that $\Delta V_\theta=\Delta V_{\rm t}$ whereas for $\theta=\pi/2$ one finds that $\Delta V_\theta=\Delta V_{\rm t}+\Delta V_{\rm e}$ as expected.

\begin{figure}
  \begin{center}
    \epsfysize=2in
    \epsfxsize=4in
    \epsfig{figure=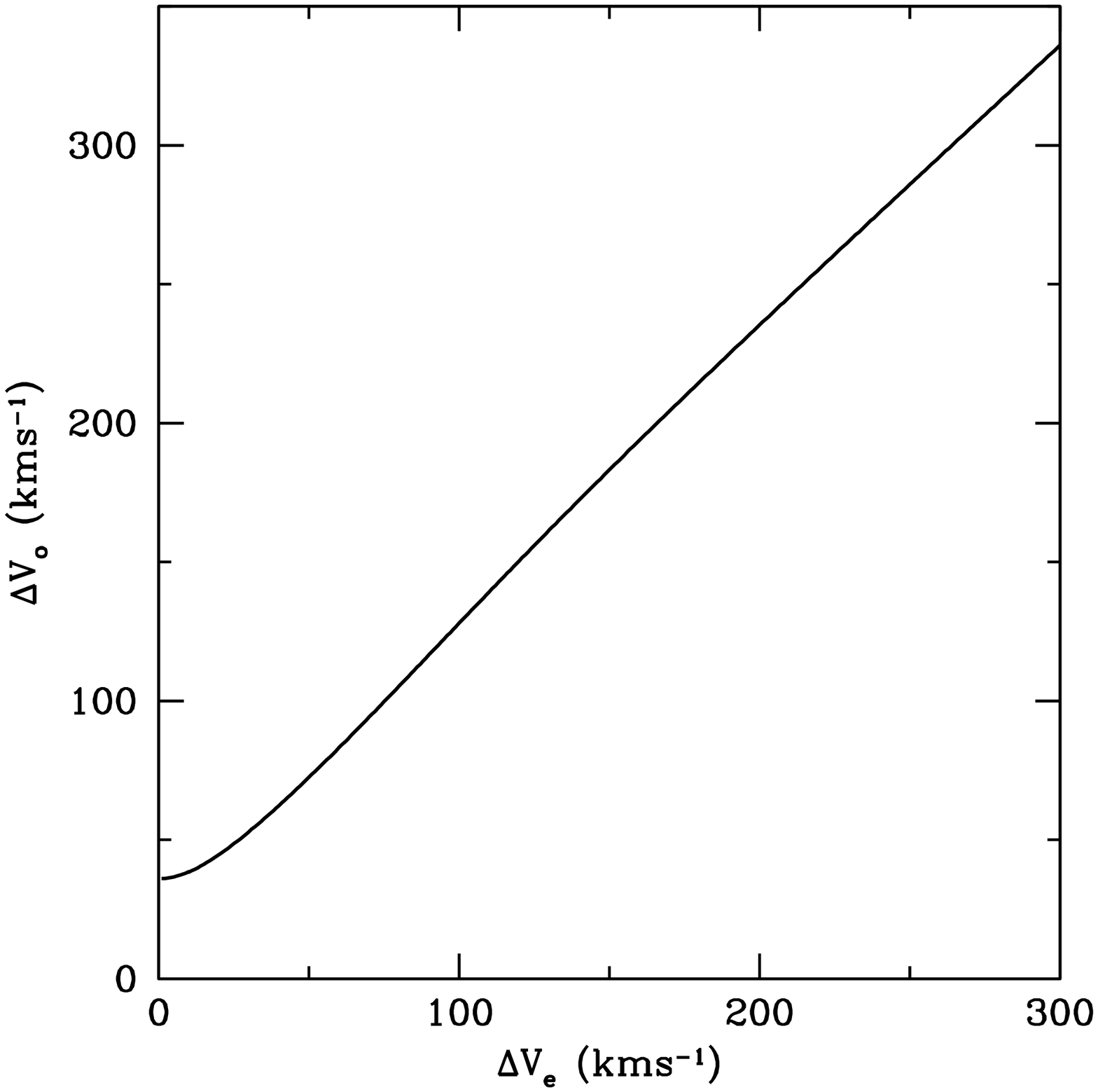, scale=0.4}
    \epsfig{figure=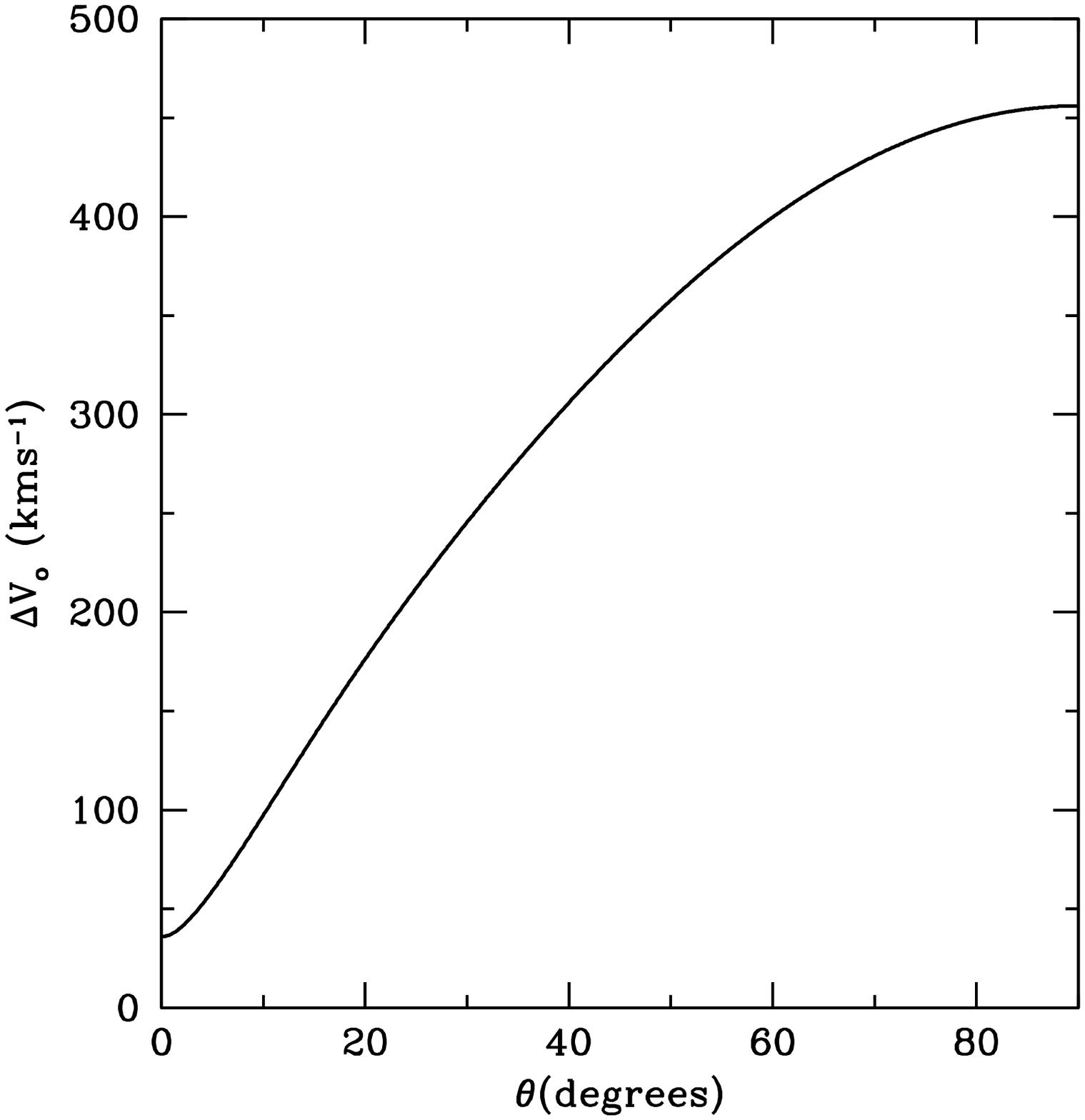, scale=0.4}
    \caption{The effective observed linewidth $\Delta V_{\rm o}$ as a function of a) $\Delta V_{\rm e}$ for $\theta=\pi/2$, that is edge on; b) $\theta$ for $\Delta V_{\rm e}=420\,{\rm km}\,s^{-1}$.}
    \label{fig:TFqplot}
  \end{center}
\end{figure}

In addition there is a broadening effect, $\Delta V_{\rm inst}$, of the HI profile due to the frequency resolution of the instrument, $R$. For a range of galaxy profiles, this broadening is found to be $\Delta V_{\rm inst}\approx 0.55R$~\citep{bot}. An appropriate value for the present discussion is $\Delta V_{\rm inst} \approx 16 \,{\rm km}\,s^{-1}$. It has been shown~\citep{hijass} that $\Delta V_{\rm inst}$ should be added linearly to $\Delta V_{\theta}$ to give the effective observed linewidth,
\begin{eqnarray} 
\Delta V_{\rm o}(\theta)= \Delta V_{\theta} + \Delta V_{\rm inst}\,.
\end{eqnarray}
$\Delta V_{\theta}$ can be computed by using the Newton-Raphson method to find the root of Eqn.~\ref{eq:TFq} for a given $\Delta V_{\rm e}(M_{\rm HI})$. 
The effective observed linewidth is illustrated in Fig.~\ref{fig:TFqplot} as a function of $\Delta V_{\rm e}$ for an edge on galaxy and as function of $\theta$ for $\Delta V_{\rm e}=420\,{\rm km}\,s^{-1}$.

\begin{figure}
  \begin{center}
    \epsfysize=2in
    \epsfxsize=4in
    \epsfig{figure=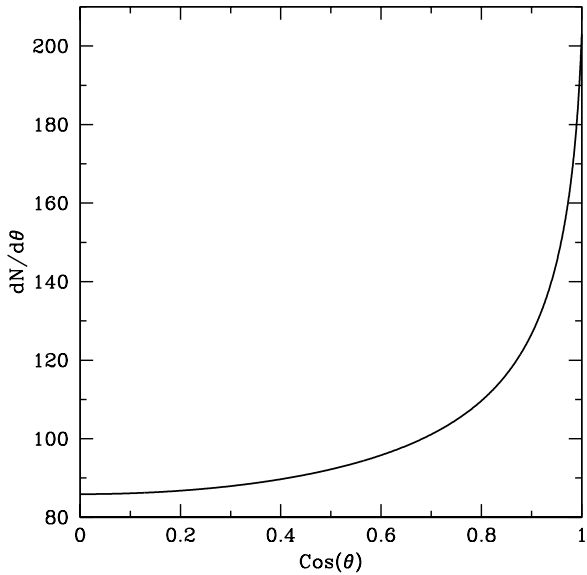, scale=0.4}
    \caption{The effect of the angle of inclination on the detectability of galaxies. We have used $z=0.1$, $\Delta z=0.01$, $t_{\rm obs}=600\,s$ and $S/N=4$.}
    \label{fig:angle}
  \end{center}
\end{figure}

The limiting mass will now depend on $\theta$ as well as $z$, $M_{\rm lim}(z,\theta)$, since it will be easier to find objects which have $\theta=0$. In order to incorporate this we have take into account the probability of a given angle $\theta$, ${\cal P}(\theta)$, when computing the number counts
\begin{eqnarray}\label{eq:improved} N \left( M > M_{\rm lim} \right) = \int^{\pi/2}_0d\theta\,{\cal P}(\theta)\,{dN\over d\theta}\,,
\end{eqnarray}
where 
\begin{eqnarray}
{dN\over d\theta}=\Delta\Omega \Delta z \frac{dV}{dzd\Omega} \int_{M_{\rm lim} (z,\theta)}^{\infty} \frac{dN}{dVdM}dM\,. \end{eqnarray} 
As an illustration, we have plotted $dN/d\theta$ against $\cos\theta$ at $z=0.1$ with $\Delta z=0.01$ for $t_{\rm obs}=600\,s$  and $S/N=4$ in Fig.~\ref{fig:angle} which shows that objects with $\theta$ close to zero are preferentially selected.

A simple assumption would be that the galaxies are randomly distributed on the celestial sphere and that the spin axes are also randomly distributed. This is not the case for satallite galaxies around their central galaxy, where a small anisotropy of the order of a few percent has been observed in SDSS data~\citep{YangBoschMo2006}, but it is likely to be a good approximation for the FAST surveys discussed here which will preferentially select large, well-separated galaxies with average separation $\approx 5 {\rm Mpc}$.
In this case ${\cal P}(\theta)=\sin\theta$ and therefore one can think of the value of $\cos\theta$ being uniformly distributed. Hence, one can deduce that
\begin{eqnarray}\label{eq:improved zeroth number count express} N \left( M > M_{\rm lim} \right) = \Delta\Omega \Delta z \frac{dV}{dzd\Omega} \int_{0}^{1} d(\cos{\theta}) \int_{M_{\rm lim} (z,\theta)}^{\infty} \frac{dN}{dVdM}. \end{eqnarray}

\begin{figure}
  \begin{center}
    \epsfysize=2in
    \epsfxsize=4in
    \epsfig{figure=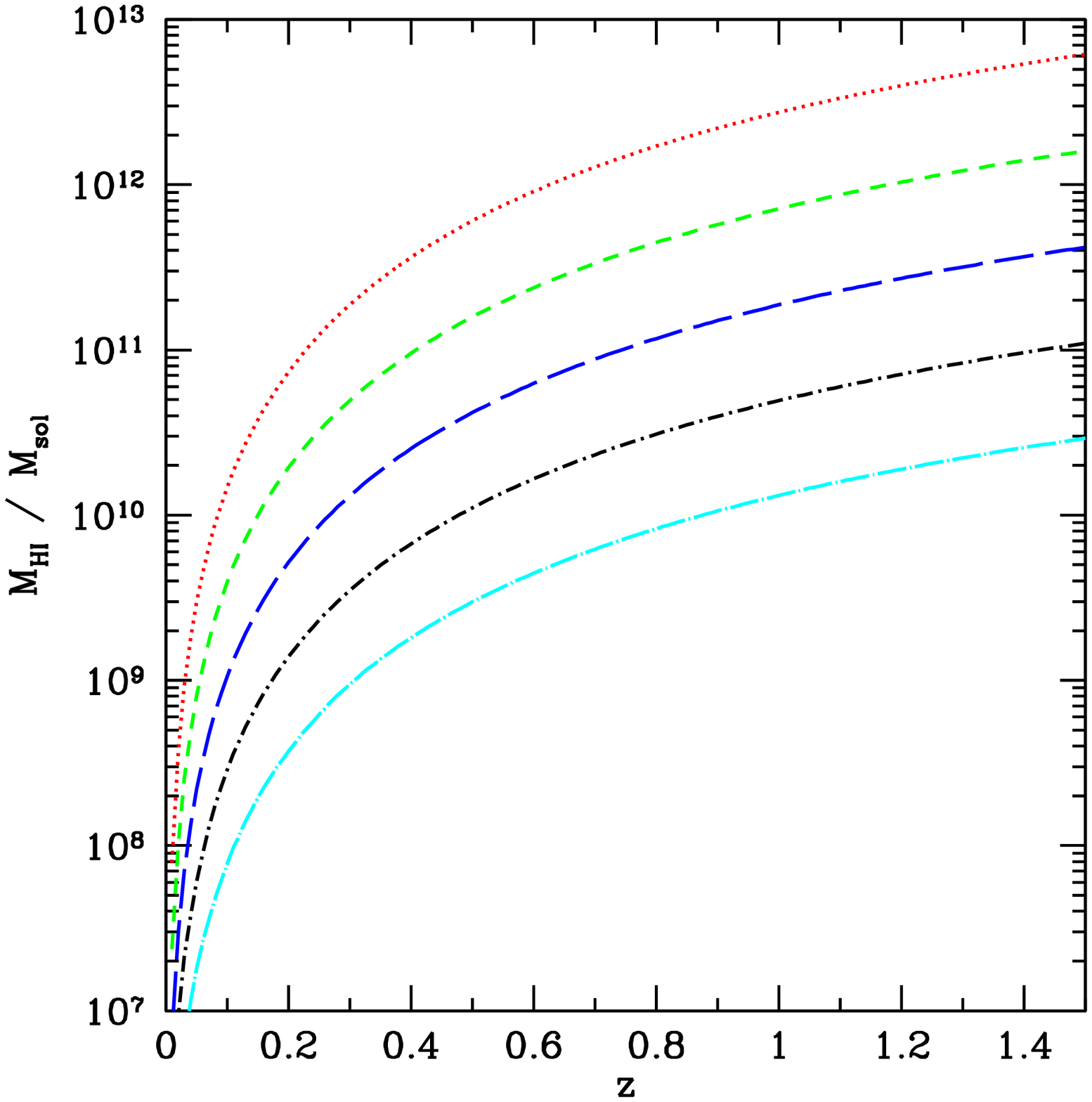, scale=0.4}
    \epsfig{figure=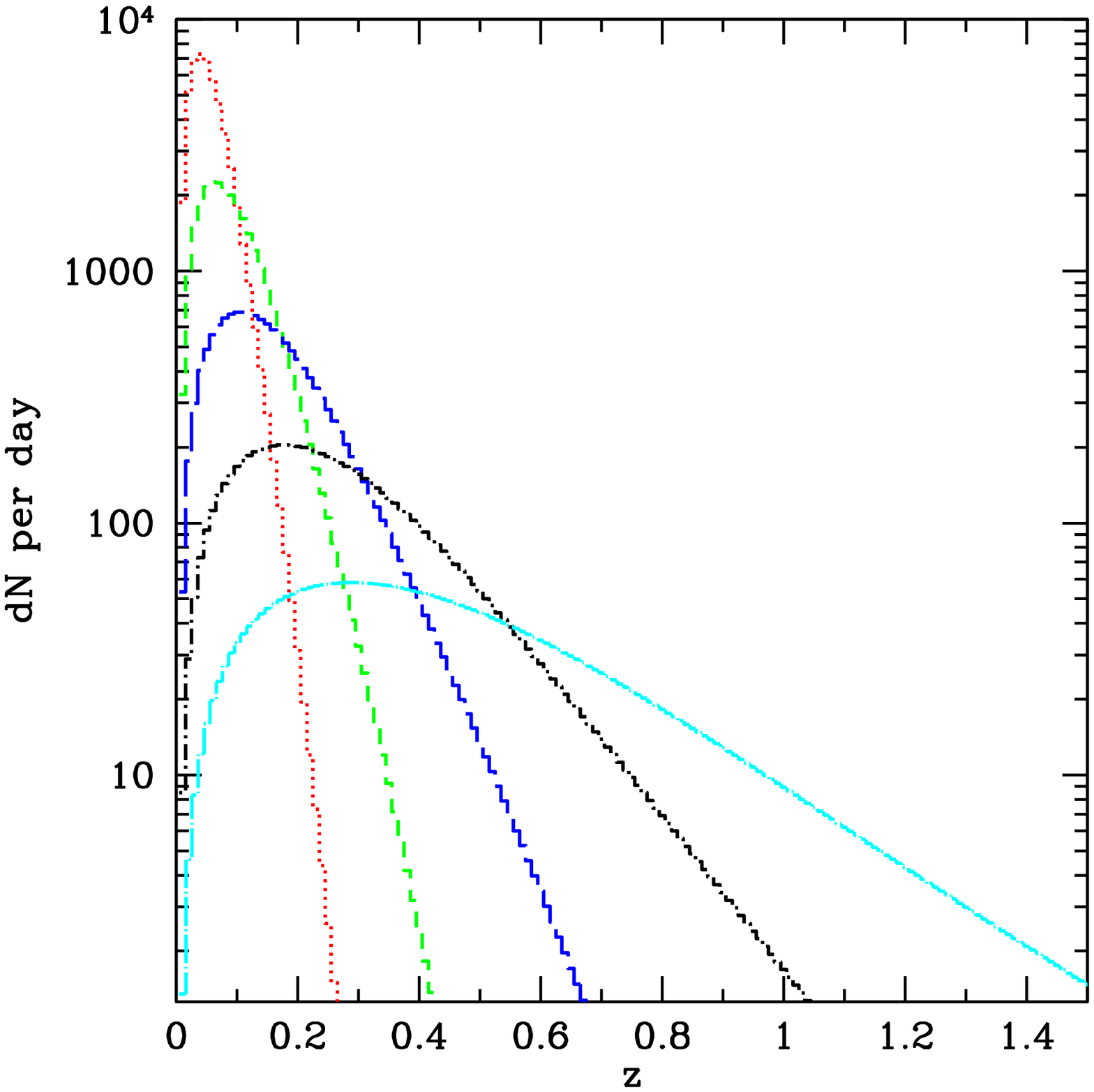, scale=0.4}
    \caption{The equivalent of Fig.~\ref{fig:simple} for the case of FAST with 100 beams when the effects of mass and inclination on the observed linewidth are taken into account.}
    \label{fig:improved}
  \end{center}
\end{figure}

The results of taking account these effects are presented in Table~\ref{tab2} and Fig.~\ref{fig:improved}. It is clear that the number counts are slightly larger, by around 30\%, than under the assumption that $\Delta V_{\rm o}=200\,{\rm km}\,s^{-1}$. The scaling formulae discussed in section~\ref{simple} apply again here. For a survey with $n_{\rm B}\approx 100$ taking around 2 years with $t_{\rm obs}=600\,{\rm secs}$ one would be able to cover all the sky available to FAST and find $\sim 10^{7}$ galaxies.

\begin{center}
\begin{table}
\begin{tabular}{|c||c||} \hline

 $t_{\rm obs}(s)$ &  $N_{4\sigma}(n_{\rm B}=19,t=1\,d)$ \\ \hline

6 & 9500  \\ \hline

60 & 5200 \\ \hline

600 & 2800 \\ \hline

6000 & 1400 \\ \hline

60000 & 690 \\ \hline

\end{tabular}
\caption{\label{tab2} The equivalent of Table~\ref{tab1} when the effects of mass and inclination on the observed linewidth are taken into account. The value of $S_{\rm lim}$ is dependent both on the mass and angle of inclination in this case, we have, therefore, not attempted to include it. All numbers have been rounded to two significant figures.}
\end{table}
\end{center}

\subsection{Effects of evolution}

The calculations of the previous section are based on the assumption that the HI mass function does not evolve. In this section we discuss the possible effects of evolution on the number counts. 
It should be noted that none of these evolving models are used in the cosmological parameter constraints in Section~\ref{error analysis}. The basic picture which we will put forward is that the amount of HI is likely to increase as we go back in cosmic time and that the estimate of number counts based on the HI mass function measured at $z=0$ is likely to be a lower limit of the number counts found by a particular survey. We will make a simple assumption that the relevant galaxies are formed by the epochs probed by FAST, and that HI is being used as the fuel for star formation, and hence is reducing in each of the galaxies as a function of time. We note that this passive star formation assumption ignores the possibility of mergers. We further note that this is only a toy model and is of course far too simplistic to be considered a realistic portrayal of the intricate physical processes acting on the HI mass function. It is designed solely as an illustrative example.

If the HI mass function is well-represented by a Schechter function then the total density of HI is given by
\begin{eqnarray}
\label{eq:HIdensity}
\rho_{\rm HI}=\Gamma(\alpha+2)\theta^*M_{\rm HI}^{*}\,.
\end{eqnarray} 
If we now make the assumption that the faint-end slope $\alpha$ does not evolve with redshift, and that $\theta^*\neq\theta^*(z)$, we can model the effective change in the hydrogen mass due to passive star formation as a net shift of the break $M_{\rm HI}^*=M_{\rm HI}^*(z)$ with redshift. We assume that a simple form of this evolution
\begin{eqnarray}
\label{eq:assumption}
M_{\rm HI}^*(z) = M_{\rm HI}^*(0) (1 + \beta z),
\end{eqnarray} 
where $\beta$ is some constant of proportionality that determines the rate at which the hydrogen mass is consumed by star formation. This should be valid at low redshifts. In the subsequent discussion we will assume that $0\le\beta\le 3$ in this model.

 One can determine values compatible with present knowledge of the evolution of $\Omega_{\rm HI}$. Under the assumption of Eqn.~\ref{eq:assumption} the total neutral hydrogen density of Eqn.~\ref{eq:HIdensity} is given by
\begin{eqnarray}
\label{eq:HIdensity2}
\rho_{\rm HI} = \Gamma(\alpha + 2) \theta^{*} M_{\rm HI}^*(0)(1 + \beta z) = \rho_{\rm HI}(z = 0) (1 + \beta z)\,.
\end{eqnarray} 

Using the cosmic density of neutral gas $\Omega_{\rm g}(z)$ (measured relative to the present day critical density) one can estimate the value of $\beta$ by assuming that the fraction of HI and He is constant between $z=0$ and $z=1$. 
By taking the ratio of measured neutral gas cosmic densities in the local Universe, $\Omega_{\rm g}(z\approx 0 )=3.5 \times 10^{-4}$, and from damped Lyman-$\alpha$ systems at higher-redshift~\citep{Prochaska2005,RaoTurnshek}, $\Omega_{\rm g}(z\approx 1)=1.0\times 10^{-3}$, we can estimate $\beta\sim 2$, which is also compatible with other works~\citep{Lah:2007nk}.

\begin{figure}
  \begin{center}
    \epsfysize=2in
    \epsfxsize=4in
    \epsfig{figure=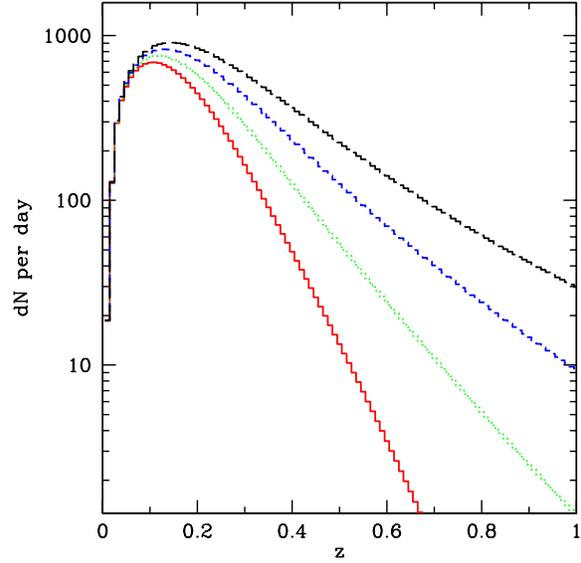, scale=0.4}
    \caption{The predicted daily number counts with $S/N=4$, $\Delta z=0.01$ binwidth, for the FAST operating with $n_{\rm B}=100$ and $t_{\rm obs}=600\,s$ as a function of redshift with a simple redshift-evolving mass break in the Schechter function of Eqn.~\ref{eq:schechterfn}. This is shown as a factor $\beta= [0,1,2,3]$ for the solid, dot, dash, dot-dashed lines respectively.}
    \label{fig:evonumber}
  \end{center}
\end{figure}

The effects of including this mass evolution for are shown in Fig.~\ref{fig:evonumber} for the 100-beam FAST with $t_{\rm obs}=600\,s$. It is clear that evolution leads to an increase in the number of galaxies found per day which are 14420, 19070, 25190, 32780 for $\beta=0,1,2,3$ and the median redshift increases with $\beta$. Essentially, we have show that one can take the current prediction of $\sim 10^{7}$ galaxies found by the 100-beam FAST in 2 years of observation is a conservative lower bound. If we take seriously the value of $\beta=2$ then one would expect to find nearly twice as many galaxies as in the case of no evolution.

\begin{figure}
  \begin{center}
    \epsfysize=2in
    \epsfxsize=4in
    \epsfig{figure=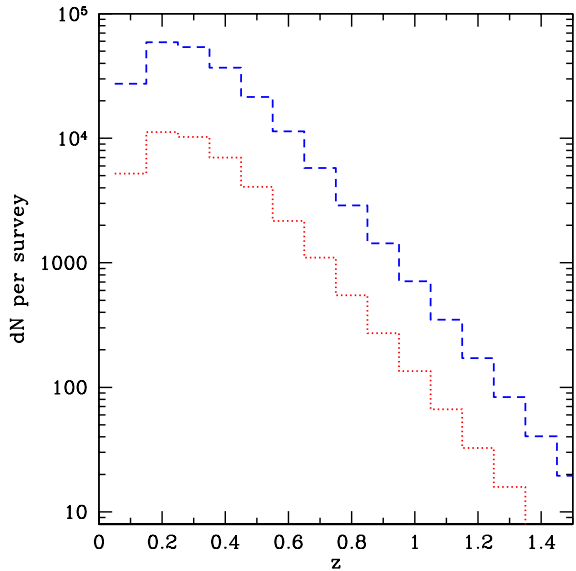, scale=0.4}
    \caption{The number of galaxies that would be found in a survey lasting $30 d$ with $t_{\rm obs}=6000\,s$. In the case of $n_{\rm B}=19$ one expect to find around 42000 galaxies with $\langle z\rangle\approx 0.3$ in $17{\rm deg}^2$, whereas one might find as many as 222000 in $87{\rm deg}^2$ for $n_{\rm B}=100$. We have used $\Delta z=0.1$ and the $z=0$ HI mass function in both cases.}
    \label{fig:deepsurvey}
  \end{center}
\end{figure}

Prior to embarking on the full, all-sky galaxy redshift survey, it would be sensible to perform an exploratory, deeper survey of a much smaller area, with the objective of measuring the evolution of the HI mass function. As well as being of legitimate scientific interest in its own right such a survey would allow one to investigate the optimal depth of the main redshift survey and pin-down some of the questions raised in this section. In Fig.~\ref{fig:deepsurvey} we show the redshift distribution of galaxies that one would expect for an HI survey with $t_{\rm obs}=6000s$ lasting $30\,d$ using the $z=0$ mass function and bins $\Delta z=0.1$. One would expected to find around 42000 galaxies with $\langle z\rangle\approx 0.3$ using $n_{\rm B}=19$ and 222000 for $n_{\rm B}=100$. In order to estimate accurately the parameters of the mass function, we estimate that one would require $\sim 1000$ galaxies per bin. Therefore, we believe that it should be possible to determine the HI mass function out to $z\approx 0.6$ for $n_{\rm B}=19$ and to $z\approx 1$ $n_{\rm B}=100$ with such a month-long survey.

\section{Expected errors on the matter power spectrum}
\label{error analysis} 
We now use the predicted galaxy number counts, assuming no evolution of the kind described in the last section, to estimate the errors of the galaxy power spectrum at $z=\langle z\rangle\approx 0.15$. $P(k,z)$ is related to the power spectrum $P(k,0)$ by \begin{eqnarray}
P(k,z)=[D(z)]^{2}P(k)\,,
 \end{eqnarray} 
where $D(z)$ is the growth factor computed from
\begin{eqnarray} D(z)=\frac{5 \Omega_{m}}{2}E(z)\int^{\infty}_{z}\frac{(1+z^{\prime})dz^{\prime}}{[E(z^{\prime})]^3}\,,
\end{eqnarray} 
an $E(z)=H(z)/H_0$.

Errors on the power spectrum are due to two factors: sample variance, i.e. the fact that not all $k$ modes are measured, and shot-noise which is the effective noise on the measurement of an individual mode. The total error $\sigma_{\rm P}$ on the measurement of the power spectrum, $P(k,z)$, for a given $k$ with logarithmic bin width $\Delta(\log k)$ can be expressed as~\citep{FKP,Tegmark1997}
\begin{eqnarray}\label{eq:Power error} \left( \frac{\sigma_{P}}{P} \right)^{2}=2\frac{1}{4 \pi k^{3} \Delta(\log k)} \frac{(2\pi)^{3}}{V_{\rm eff}(k)} \left( \frac{1+nP}{nP} \right)^{2}\,,\end{eqnarray}
where $P=P(k,z)$ and $n=n(z)$ is the number density of galaxies which are detected (making $nP$ dimensionless)
\begin{eqnarray}
n(z)=\int_{M_{\rm lim}(z)}^{\infty}{dN\over dVdM}\,dM\,,
\end{eqnarray}
and $V_{\rm eff}(k)$ is the effective survey volume probed for a particular $k$-mode 
\begin{eqnarray}
\label{eq:veff}
V_{\rm eff}(k)=\Delta \Omega \int_{0}^{\infty} \left(\frac{nP}{1+nP}\right)^{2} \frac{dV}{dzd\Omega}(z) dz\,. 
\end{eqnarray}

For definiteness, we will consider a survey taking two years performed by the FAST telescope with a focal plane array with $n_{\rm B}=100$ and $t_{\rm obs}=600\,s$ per instantaneous FoV, $\Omega_{\rm inst}\approx 1000\,{\rm arcmin}^2$. The choice of $t_{\rm obs}=600\,s$ is simply the required survey time for scanning the full sky in 2 years and hence is the most efficient use of the telescopes resources. The justification for such an observation strategy is the simple fact that the area covered is inversely linear in integration time per field while the depth attained goes as the square root of the integration time; the most cosmic volume is attained from full sky surveys. We have also calculated the effect of probing different baryonic acoustic modes in the sky with varying observation strategies and determined that the simple maxim of `wide-shallow' is always preferable to `narrow-deep' in large-scale galaxy surveys. Using the methods of section~\ref{inclination} we have estimated that such a survey will find  $\sim  10^{7}$ galaxies with $S/N=4$, the vast majority of which are spatially unresolved.

\begin{figure}
  \begin{center}
    \epsfysize=2in
    \epsfxsize=4in
    \epsfig{figure=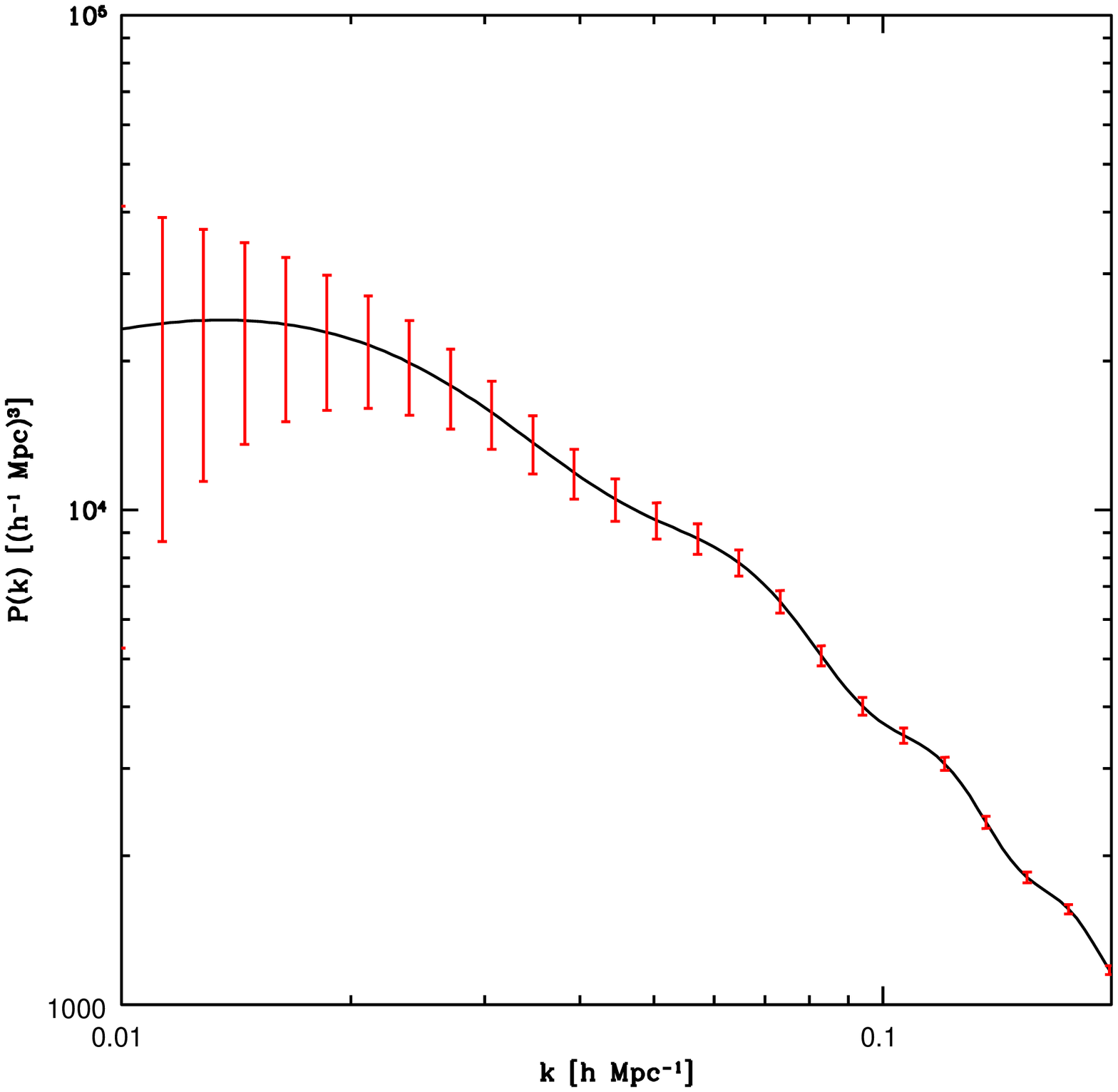, scale=0.4}
    \epsfig{figure=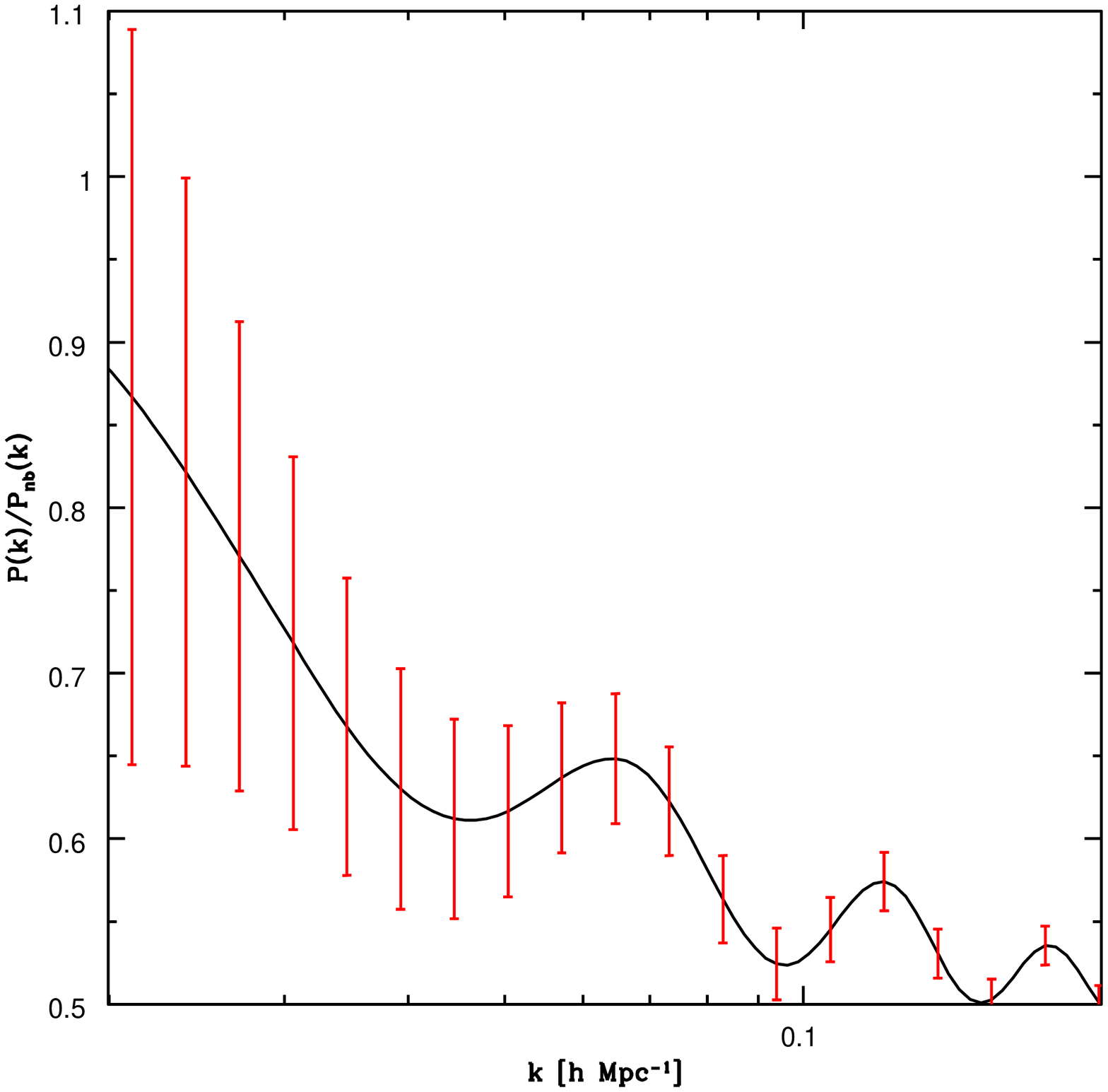, scale=0.4}
    \caption{a) the expected errors on the power spectrum for the proposed FAST survey; b) the same errors, but on a scale where the power spectrum has been divided by the equivalent no baryon model.}
    \label{fig:powersp}
  \end{center}
\end{figure}

The expected errors on the power spectrum are presented in Fig.~\ref{fig:powersp} for $\Delta(\log k)=0.04$, which should be sufficiently large to ensure the errors are uncorrelated, according to a simple prescription detailed in~\cite{MeiskinWhitePeacock1999}.
The errors increase at low $k$ due to sample variance and would also increase at high $k$ due to shot noise were it not for our choice of a logarithmic binning; that is, the bins do not contain an equal number of $k$-modes. We have also divided the power spectrum by a model with no baryons which illustrates the acoustic features in the spectrum. The average redshift $\langle z\rangle\approx 0.15$ will prevent much cosmological information being gleaned from  the acoustic scale and therefore we will not use it explicitly in our estimates of the cosmological parameters. 
It is clear, however, from Fig.~\ref{fig:powersp} that the detection of the baryonic features would be possible with these data.

\begin{figure}
  \begin{center}
    \epsfysize=2in
    \epsfxsize=4in
    \epsfig{figure=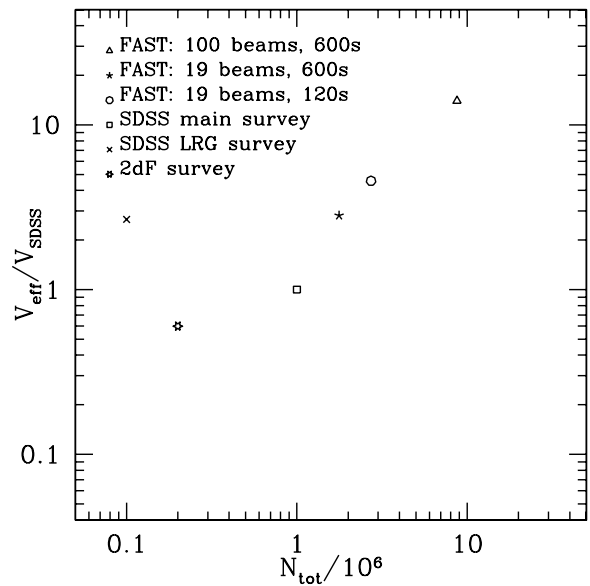,scale=0.4}
    \caption{The effective $V_{\rm eff}$ at $k=0.075h\,{\rm Mpc}^{-1}$ for various completed and proposed surveys plotted against the approximate number of galaxies they have or might find. The volumes are quoted in terms of that of the SDSS main galaxy sample which has $V_{\rm SDSS}=1.5\times 10^{8}h^{-3}\,{\rm Mpc}^3$. The open square is for the SDSS main sample, the cross is for the SDSS LRG sample, the hexagon is for the 2dFGRS, the pentagon is FAST with $n_{\rm B}=19$ with $t_{\rm obs}=600\,s$ while the circle is with $t_{\rm obs}=120\,s$ and the open triangle is for the proposed FAST survey taking 2 years with $n_{\rm B}=100$.}
    \label{fig:effvol}
  \end{center}
\end{figure}

It is now interesting to compare the characteristics of the proposed FAST survey with those presently available. The effective volume of the proposed FAST survey evaluated at the $k=k_{\rm A}=0.075h\,{\rm Mpc}^{-1}$ scale of the acoustic peak in the power spectrum, is $V_{\rm eff}=0.82\times 10^{9} h^{-3}\,{\rm Mpc}^{3}$.  This is shown in Fig.~\ref{fig:effvol} where the effective volume probed, as well as the galaxy number count, of a variety of optical survey schemes are given, namely the SDSS main and SDSS LRG results as well as 2dFGRS, in comparison with FAST $n_{\rm B}=19$ for both $t_{\rm obs}=120\,s$ and $t_{\rm obs}=600\,s$ survey modes as well as our proposed version with $n_{\rm B}=100$ and $t_{\rm obs}=600\,s$.

\section{Error forecasts for cosmological parameters}
\label{sec:errorforecasts}

\begin{center}
\begin{table*}
\begin{tabular}{|c||c|c||c|c|c||} \hline
 Parameter                       &  SDSS           &  2dFGRS       & FAST (19/120) & FAST (19/600)   & FAST (100/600)  \\ \hline
$\delta\Omega_{\rm m}/\Omega_{\rm m} $ & $0.416$ & 0.221 & 0.230 & 0.310   & 0.078   \\ \hline
$\delta\Gamma/\Gamma$ & $0.328$ & 0.161 & 0.166 & 0.231   & 0.049   \\ \hline
$\delta f_{\rm b}/f_{\rm b}$ & $0.378$ & 0.253 & 0.468 & 0.486   & 0.283   \\ \hline
$\delta n_{\rm s}/n_{\rm s}$ & $0.251$ & 0.141 & 0.148 & 0.178   & 0.071   \\ \hline
\end{tabular}
\caption{\label{tab:fastsummary} Summary of parameter values which come from the SDSS main sample and 2dFGRS, and the simulated constraints expected from the proposed FAST surveys with : (i) $n_{\rm B}=100$, $t_{\rm obs}=600\,s$; (ii) $n_{\rm B}=19$, $t_{\rm obs}=120\,s$; (i) $n_{\rm B}=19$, $t_{\rm obs}=600\,s$. Values presented are the fractional errors for each parameter.}
\end{table*}
\end{center}
In the previous section we have shown that a survey using the FAST with $n_{\rm B}=100$ would find $\approx 10^{7}$ galaxies. It is interesting to determine the improvement that this would bring to cosmological parameter estimation through the accurate sampling of the galactic power spectrum relative to current large-scale optical surveys; 2dFGRS and SDSS. Since the FAST survey will have similar redshift coverage, but in a different waveband, this would bring more than just an improvement in signal-to-noise.

For the 2dFGRS survey we use the data presented in~\citep{Cole}. We select 32 bandpowers in the range $0.022 < k/{h\,\rm (Mpc^{-1})} < 0.147$, where non-linear effects are expected to be small in this region. Even so, in ref.~\citep{Cole} it was proposed to use a fitting function of the form
\begin{eqnarray} \label{eqn:2df}
P_{\rm obs} (k)=b^{2} \frac{1+Qk^{2}}{1+Ak} P_{\rm lin} (k)\,,
\end{eqnarray}
to model the residual non-linear effects on the original, linear, power spectrum $P_{\rm lin}(k)$. For the 2dFGRS data, the authors advocate using the parameters $A = 1.4\,{\rm Mpc^{-1}}$ and $Q = 4\,{\rm Mpc^{-2}}$. We follow this treatment, and analytically marginalize over the bias parameter $b$ in the subsequent Markov-Chain Monte-Carlo (MCMC) analysis. For the SDSS data, we use data presented in ref.~\citep{Tegmark2004a}, using 19 bandpowers in the range $0.016 < k/(h{\rm  Mpc^{-1}}) < 0.205$.  We take the power spectrum of galaxies to be of the form $P_{\rm obs} (k)=b^{2}P_{\rm lin} (k)\,,$ and analytically marginalize over the bias. For each of these datasets we use the window functions provided by~\citep{Cole} and~\citep{Tegmark2004a} respectively. 

For the FAST data, we generated a fiducial power spectrum using the {\tt CAMB} software~\citep{Lewis:1999bs} using cosmological parameters that best-fit the $\Lambda$CDM model~\citep{WMAP3}, in this case [$\Omega_{\rm c} h^2$, $\Omega_{\rm b} h^{2}$, $h$, $n_{\rm s}$, $A_{\rm s}$] are given by [0.104, 0.0223, 0.734, 0.951, $2.02 \times 10^{-9}$], where $A_{\rm s}$ is the initial amplitude of fluctuations. We then sample $P_{\rm lin} (k)$ at closely spaced values of $k/{\rm h}$, from which we interpolate to obtain $P_{\rm lin} (k)$ at the appropriate FAST survey $k/{\rm h}$ values, along with the simulated error bars. 

For completeness we have investigated three FAST survey schemes, namely the fiducial 19-beam FAST, with integration times per pixel of $t_{\rm obs}=120\,s$ and 600 secs, and the 100-beam system with 600 seconds of integration per pixel each of which would take $\sim 2$ years to complete. For each scheme we use 28 bandpowers in the range $0.005 < k/{\rm (h Mpc^{-1})} < 0.155$, and take the galaxy power spectrum to have the form $P_{\rm obs} (k)=b^{2} P_{\rm lin} (k)\,$. We also assume that the errors on each of the bins are uncorrelated due to the large binwidths taken in $k$ as mentioned in Section~\ref{error analysis}. The 19-beam survey with $t_{\rm obs}=120\,s$, which could find $\approx 3\times 10^{6}$, galaxies would cover the full $20000\,{\rm deg}^2$, albeit to a shallower depth than the 100-beam case, whereas the 19-beam $t_{\rm obs}=600\,s$ survey would only cover $4000\,{\rm deg^2}$ and find $\approx 2\times 10^{6}$ galaxies. 

For the MCMC analysis we use {\tt COSMOMC}~\citep{Lewis:2002ah} to create chains to estimate the confidence limits on the cosmological parameters. Since large parameter degeneracies exist when only including large-scale structure data in the fit, we impose the consistency relation that the angle $\theta_{\rm acoustic}$ subtended by the first acoustic peak is $1.040$, which is strongly constrained by the WMAP3 data~\citep{WMAP3}.  This essentially leaves three remaining free parameters in the fit: $\Omega_{\rm c} h^2$; $\Omega_{\rm b} h^{2}$ and $n_{\rm s}$, which are used to compute the parameter $\Gamma=\Omega_{\rm m}h$ and $f_{\rm b}=\Omega_{\rm b}/\Omega_{\rm m}$. This is because we  marginalize over bias, which absorbs the initial amplitude of fluctuations, and $h$ can then be derived from these parameters and $\theta_{\rm acoustic}$.

In Table~\ref{tab:fastsummary} we present the fractional error on each of the marginalized cosmological parameters for the 2dFGRS and SDSS datasets, along with the three FAST observing schemes. The errorbars on $f_{\rm b}$ are of a similarly poor level to those possible with 2dFGRS and SDSS. This is probably since dependence of the power spectrum on $f_{\rm b}$ is weak and the present surveys have reached the ceiling on how well this can be measured from the matter power spectrum. However, the errors on $\Gamma$ and $n_{\rm s}$ are significantly improved for the case of $n_{\rm B}=100$ and  $t_{\rm obs}=600\,s$, the area of the error ellipse in the $\Gamma-n_{\rm s}$ direction is reduced by about a factor of six from that possible with 2dFGRS. 
It is clear, however, that the FAST surveys with $n_{\rm B}=19$ do not significantly improve on the constraints already available. Of the two, that with $t_{\rm obs}=120s$ which covers the whole sky does much better than $t_{\rm obs}=600\,s$ making it clear, yet again, that surveys which find the largest number of objects (and which cover all the available sky) will typically constrain cosmological parameters the best.

\begin{figure*}
  \begin{center}
    \epsfysize=2in
    \epsfxsize=4in
    \epsfig{figure=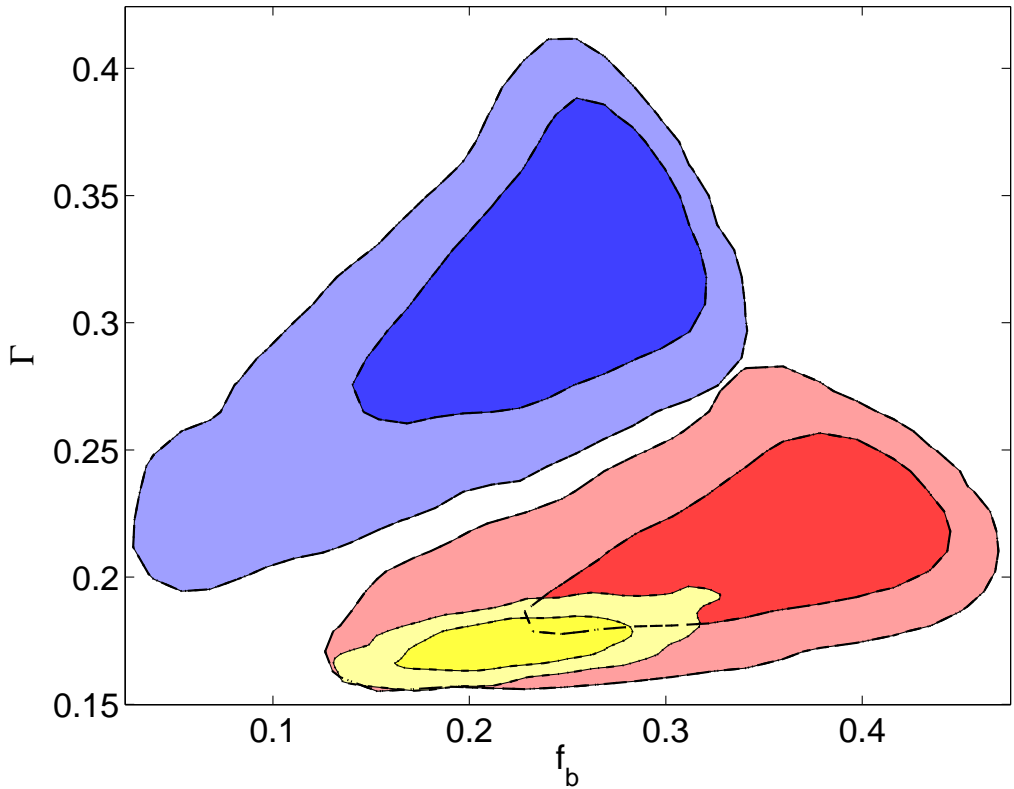,scale=0.8}
    \caption{Cosmological parameter estimation for three surveys: the SDSS (top, blue); 2dFGRS (middle, red) and the proposed 100-beam FAST (bottom, yellow). Contours are $1\sigma$ and $2\sigma$ respectively representing the $68\%$ and $95\%$ probability estimates. Due to the tension between the 2dFGRS and SDSS central values, the fiducial FAST cosmology was based on WMAP3. Note the large improvement in constraining this parameter space thanks to the order of magnitude increase in the number counts as well as in the volume probed.}
    \label{fig:constraints}
  \end{center}
\end{figure*}

In Fig.~\ref{fig:constraints} we present marginalized 2D likelihoods of $\Gamma$ versus $f_{\rm b}$ . Here, we impose a prior of $n_{\rm s}=0.95 \pm 0.02$, as preferred by the WMAP3 data~\citep{WMAP3}. The first thing that is apparent is that the SDSS and 2dFGRS do not agree on the central value; something which has been noted in the literature and is thought to be associated with the different selection criteria for the two surveys~\citep{Cole:2006kn}. What is also evident is that FAST with $n_{\rm B}=100$ performs considerably better than the two presently available surveys.

Low redshift galaxy redshift surveys cannot constrain the properties of dark energy directly, but they can in combination with measurements of the CMB. Essentially, the measurement of $\Gamma$ by the redshift survey breaks the angular diameter degeneracy, allowing simultaneous measurements of $h$ and the equation-of-state parameter of the dark energy $w=P/\rho$. In order to see how FAST could impact on this we have performed a joint analysis using a simulated Planck dataset using a noise power spectrum for the temperature anisotropies out to $\ell_{\rm max}=2000$ of $N_{\ell}^{\rm TT}=9\times 10^{-5}\mu{\rm K}^2$, and that for the E-mode polarization given by $N_{\ell}^{\rm EE}=4N_{\ell}^{\rm TT}$. We have assumed full sky observations and that optimal uncorrelated errors can be reconstructed from the data. The results are presented in table~\ref{tab:planck} which show that $w$ can be measured to within $5\%$ of -1, and $h$ to within $\approx 2\%$.
 
\begin{center}
\begin{table}
\begin{tabular}{|c||c|c||} \hline
 Parameter &  Planck &  Planck + FAST 100/600 \\ \hline
$\Omega_{\rm b}h^2 $  & $0.0223\pm 0.0002$ & $0.0223\pm 0.0002$  \\ \hline
$\Omega_{\rm c}h^2 $  & $0.104\pm 0.002$ & $0.104\pm 0.002$  \\ \hline
$n_{\rm s}$  & $0.952\pm 0.005$ & $0.952\pm 0.005$  \\ \hline
$\log(10^{10}A_{\rm s})$  & $3.01\pm 0.01$ & $3.01\pm 0.01$ \\ \hline
$h$  & $0.751\pm 0.131$  & $0.731\pm 0.015$  \\ \hline
$\tau$  & $0.090\pm 0.006$ & $0.090\pm 0.0006$  \\ \hline
$w$ & $-1.02 \pm 0.328$ & $-0.99\pm 0.05$ \\ \hline
\end{tabular}
\caption{\label{tab:planck} Expected errors for a 7 parameter fit using Planck alone and Planck + FAST 100/600. There is a degeneracy between $w$ and $h$ for Planck only, which is broken by the inclusion of FAST. $\tau$ is the optical depth to reionization, asssumed to be instantaneous, and $A_{\rm s}$ is the standard power spectrum amplitude.}
\end{table}
\end{center}

\section{Comparison with early-phase SKA}

An early phase of the SKA with, say, $10\%$ of the total collecting area (that is, an effective collecting area of $\sim 70,000 {\rm m}^2$ and thus $\sim 40 \%$ larger than FAST) and a FoV of at least $\Omega_{\rm inst}=1\,{\rm deg}^2$ (that is more than three times larger than the 100-beam case) is likely to come into operation during the first five years of FAST operation. It is interesting to compare and contrast the likely performance of such an early-stage SKA relative to FAST. We will assume that, for HI observations, the value of $T_{\rm sys}$ for the 10\% SKA is the same as for FAST (we note that it is likely to be slightly higher, but not appreciably so; our results can be scaled by $T_{\rm sys}^{-2}$ where appropriate). The angular resolution will, however, be much higher since the baselines on the SKA ``core'' will extend out to $\sim 5$ km and hence the synthesised beam will be $\sim 10$ arcsec. This introduces an extra complication into predicting the number counts since many galaxies may be resolved and hence use of the point source approximation will overestimate the number of objects detected, particularly for shallow surveys. 

In order to model this effect we have introduced an upper limit in the integral in eqn.~\ref{eq:zeroth} which is defined as the mass of object at a specific redshift which is resolved by the telescope beam. Of course, this then underestimates the number of objects found since some resolved objects are always found in any survey. However, the number which are missed by this procedure are likely to be small since the vast majority of detections are just above mass threshold.

\begin{figure*}
  \begin{center}
    \epsfysize=2in
    \epsfxsize=4in
    \epsfig{figure=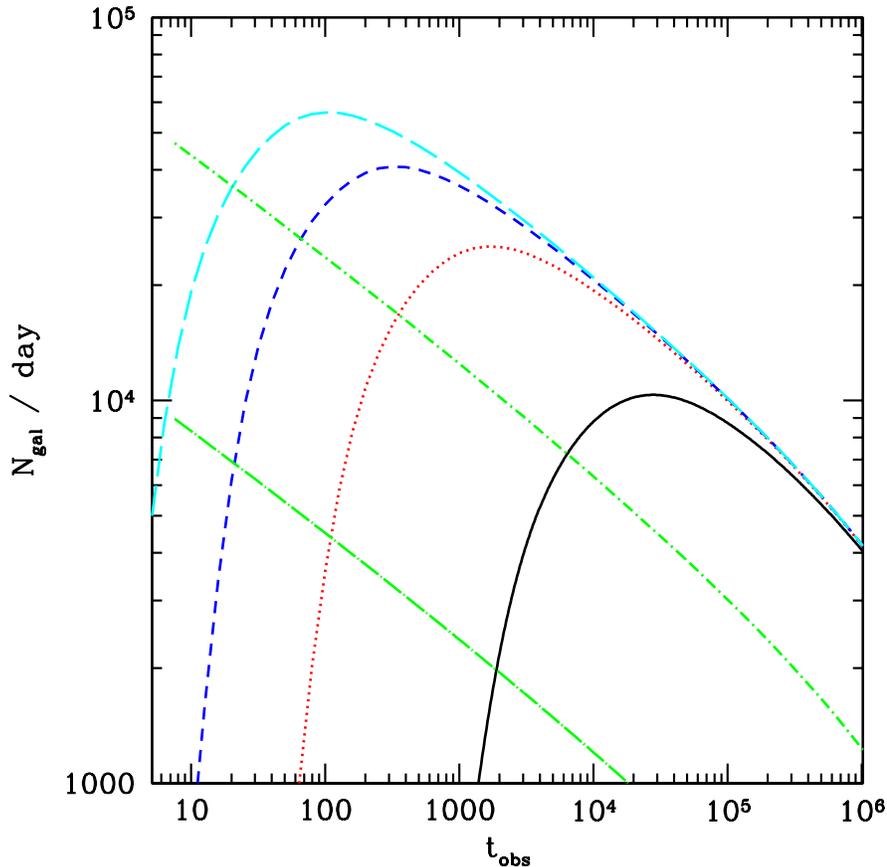,scale=0.6}
    \caption{$10\%$ SKA number counts ($6\sigma$) for a variety of surveying strategies defined by $t_{\rm obs}$, with a range of proposed angular resolutions. The resolutions considered here are 6 arcsec (solid), 12 arcsec (dot), 18 arcsec (dash) and 24 arcsec (large-dash). The 19-beam (dotted long-dash) and 100-beam (dotted short-dash) surveys possible with FAST are presented for comparison. As explained in the text the $10\%$ SKA will be significantly limited by the effect of resolving out galaxies, hence the trend seen here of better resolution demanding a survey scheme that favours deeper exposures to probe higher redshifts and observe more galaxies as point-like.}
    \label{fig:tenpercent}
  \end{center}
\end{figure*}

The number of detections expected in one day are presented in Fig.~\ref{fig:tenpercent} for a range of angular resolutions between 6 and 24 arcsec as a function of $t_{\rm obs}$. Included also for comparison are predictions for FAST with 19 and 100 beams. The first thing to notice is that the FAST curves are almost straight lines which peak at the lowest possible value of $t_{\rm obs}$ confirming our earlier assertion that the best observing strategy for FAST, in terms of the number of objects detected, is to choose the value of $t_{\rm obs}$ so that one would cover the whole sky available to FAST in a defined survey time. The curves for the 10\% SKA have a very different shape, following a line similar to that for the FAST case for large values of $t_{\rm obs}$ before eventually turning over and plumetting to zero. This is the effect of the finite resolution, with the maximum being where the majority of objects found in a survey fill the beam exactly.

It is clear to see that for $\theta_{\rm FWHM}>15\,{\rm arcsec}$ the 10\% SKA would make most of its detections in the point source regime for $t_{\rm obs}>100\,s$. A higher resolution is more likely and if we consider the case of 12 arcsec resolution then the optimum value of $t_{\rm obs}\approx 1000\,s$, but in terms of the number of objects found per day it is only marginally more powerful than for FAST with $n_{\rm B}=100$. For higher resolution still the optimal value of $t_{\rm obs}$ increases to around $2\times 10^{4}\,s\approx 6\,{\rm hours}$ at 6 arcsec. In this case, one would  be performing a much deeper survey which might capable of probing power spectrum at much higher redshifts and constraining the dark energy using the Baryonic Acoustic Oscillations. It would certainly be very different in nature to that which we have proposed for FAST~\citep{AR}.

To summarize if the 10\% stage of the SKA has relatively low resolution and a FoV of $\sim 1$ deg$^2$ then it will be a factor of a few times more powerful than FAST, in terms of the number of galaxies found at relatively low redshifts. But if it has much higher angular resolution then the survey which it will yield will be dominated by objects at much higher redshifts. If this were to be the case the FAST and 10\% SKA would be complementary.

\section{Conclusions}

In this paper we have shown that the upcoming FAST is an extremely capable instrument for large-scale neutral hydrogen surveys over the whole sky with the potential for large numbers of galaxies  being discovered daily in the redshift range $0-0.2$ and tens of galaxies out to mid-redshifts of order $0.2 - 0.4$. This would provide complementary information to the redshift surveys already performed at optical wavelengths and would be considerably better than those presently available (or likely to be available in the near future) selected using neutral hydrogen as the tracer.

In calculating the number of galaxies expected we have taken into account the effect of different angles of inclination for the galaxies to the observer and shown that an increase of $\sim 30\%$ is expected over simplistic approaches. Essentially the increase arises from the detection of face-on galaxies which are otherwise neglected. A first-order estimate has also been made of the effect of the evolution of the  hydrogen mass of a galaxy by means of  a toy model based on a linear increase in the mass of $M^*$ with redshift. Although simplistic this nevertheless indicates that quite conservative increases in the neutral hydrogen content of a galaxy with redshift can lead to a factor of two increase in the expected number counts. This would lead to reductions in errors of the power spectrum due to shot noise, although we have not used this evolving HI mass function for the cosmological parameter constraints of Section~\ref{sec:errorforecasts}. 
Prior to the main FAST survey we have suggested a deep, month-long survey to investigate these evolution effects by estimating the HI mass function out to $z\approx 0.6$.

For the FAST HI galaxy survey to be a major improvement over current optical surveys it will be necessary to detect  $\sim 10^{7}$ galaxies over $\sim 20000$ deg$^2$, most of which will lie within redshifts of a few tenths. A uniformly sampled $\sim 10^7$ HI galaxy survey will allow nearly an order of magnitude improvement of cosmological parameter constraints than currently provided by SDSS. For example the FAST survey will, independent of CMB observations, constrain $n_s$ to 7\% and $\Gamma$ to 5\%. 
Used in conjunction with observations of the CMB made by Planck this could reduce the volume of the overall cosmological parameter space.

We have shown, however, that to achieve this performance in a reasonable time ($\sim 2$ years) a receiving system providing of order 100 beams is necessary. Conventional horn-based systems with physical limits to their packing density are restricted to $\sim 19$ beams at the FAST prime focus and hence the 100-beam requirement can only be met with close-packed phased arrays of small antenna elements; these would not take up much more area than the 19-beam horn system. The required phased array technology is actively being developed by several groups working on technology for the SKA and hence we can be confident that by the time FAST comes into operation in 2012-2013 the 100-beam receiver capability will be available.  
 
We have briefly looked ahead to the first phase of the SKA and compared the shallower FAST surveys with those from the much higher resolution interferometric surveys which will come somewhat later than that from FAST. The FAST and early-SKA surveys are complementary with the higher resolution SKA surveys preferentially finding higher redshift objects. 

\section*{Acknowledgements}

ARD is supported by a STFC studentship.  We would like to thank Nan Rendong and Bo Peng for helpful comments.

\label{lastpage}
\end{document}